\documentclass[runningheads,a4paper]{llncs}
\usepackage[top=2.5cm, bottom=2.5cm, left=2.5cm, right=2.5cm]{geometry}
\usepackage{amsmath}
\usepackage{amssymb}
\usepackage{graphicx}
\usepackage{booktabs}
\usepackage{listings}
\usepackage{xcolor}
\usepackage{hyperref}
\usepackage{caption}
\usepackage{subcaption}

\usepackage{url}
\usepackage{color}
\usepackage{cite}
\usepackage{epsfig}

\usepackage[nomargin,inline,marginclue,draft]{fixme}
\usepackage{cleveref}
\fxsetup{status=draft}
\fxsetup{theme=color, mode=multiuser} \FXRegisterAuthor{me}{ame}{\color{red}Me}
\newcommand{\orcid}[1]{\href{https://orcid.org/#1}{\includegraphics[scale=0.02]{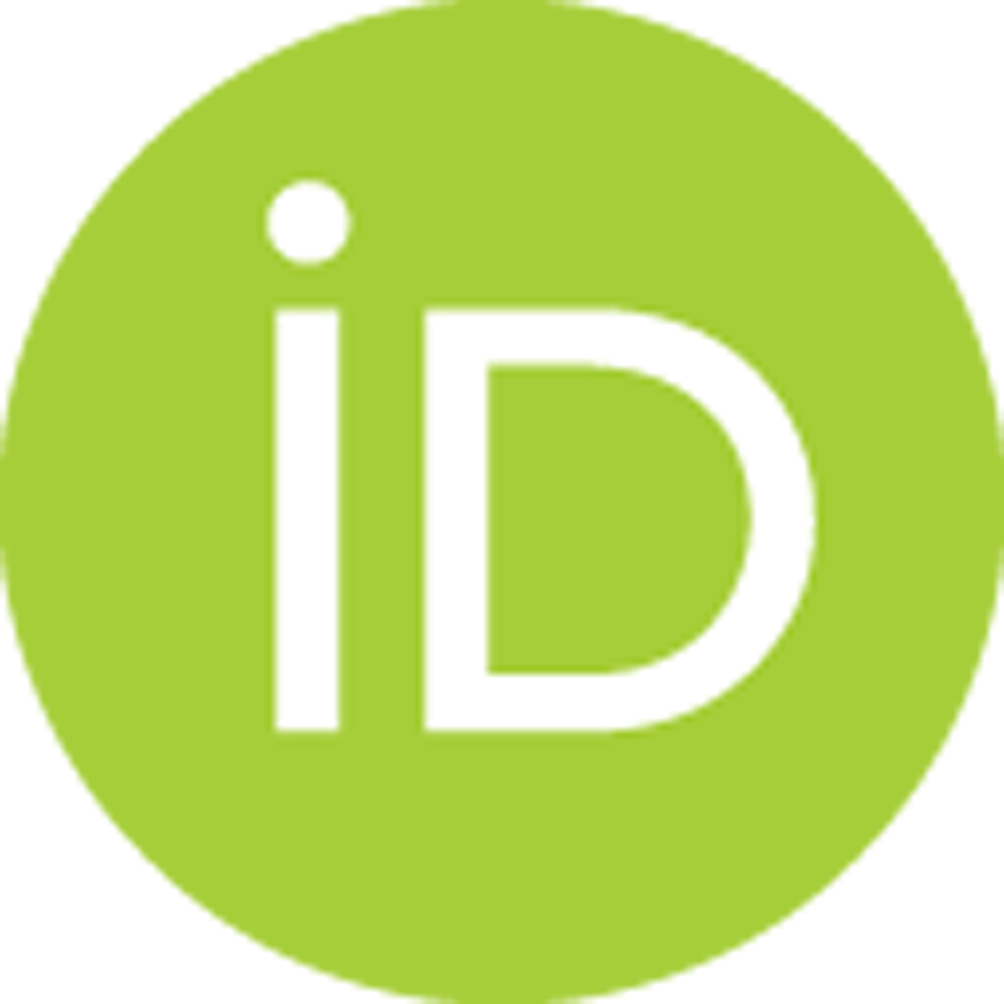}}} % this does not print the whole orcid number

\hypersetup{
    colorlinks=true
}
\title{Multiscale Encoder and Omni-Dimensional Dynamic Convolution Enrichment in nnU-Net for Brain Tumor Segmentation}
\titlerunning{ }
\author{
Sahaj K. Mistry \inst{1,**}
\and Aayush Gupta \inst{1,*}
\and Sourav Saini \inst{1,*}
\and Aashray Gupta\inst{1,*}
\and Sunny Rai \inst{2}
\and\\ Vinit Jakhetiya \inst{1}
\and Ujjwal Baid \inst{3\dagger}
\and Sharath Chandra Guntuku \inst{2\dagger}
}
\authorrunning{Sahaj, et al.}

\institute{%\scriptsize{
Indian Institute of Technology Jammu, India
\and University of Pennsylvania, PA, USA
\and Indiana University, IN, USA
\linebreak
\\
\textsuperscript{*} Equal Contributors.\\
\textsuperscript{$\dagger$} Senior authors.\\ 
\textsuperscript{**} Corresponding author: \email\{sahajmistry005@gmail.com\}}

\begin{document}
    \mainmatter
    \maketitle
    
    \bibliographystyle{plain}  % or any other style you prefer
    % \bibliography{bibliography.bib}
    % \input{main.bbl}
    
    \setcounter{footnote}{0} 
    \begin{abstract}
        Brain tumor segmentation plays a crucial role in computer-aided diagnosis. This study introduces a novel segmentation algorithm utilizing a modified nnU-Net architecture. Within the nnU-Net architecture's encoder section, we enhance conventional convolution layers by incorporating omni-dimensional dynamic convolution layers, resulting in improved feature representation. Simultaneously, we propose a multi-scale attention strategy that harnesses contemporary insights from various scales. Our model's efficacy is demonstrated on diverse datasets from the BraTS-2023 challenge. Integrating omni-dimensional dynamic convolution (ODConv) layers and multi-scale features yields substantial improvement in the nnU-Net architecture's performance across multiple tumor segmentation datasets. Remarkably, our proposed model attains good accuracy during validation for the BraTS Africa dataset. The ODconv source code\footnote{\url{https://github.com/i-sahajmistry/nnUNet_BraTS2023/blob/master/nnunet/ODConv.py}} along with full training code\footnote{\url{https://github.com/i-sahajmistry/nnUNet_BraTS2023}} is available on GitHub.
    \end{abstract}
    
    \keywords{BraTS-2023, deep learning, brain, segmentation, Adult Glioma, BraTS-Africa, Meningioma, Brain Metastases and Pediatric Tumors, nnU-Net, ODConv3D, multiscale, lesion, medical imaging}
    
    % =========================
    % main part of the document
    % =========================
    \section{Introduction}

% Glioblastomas (GBM) are the most common and aggressive malignant primary brain tumors in human adults. High morphological
% and histological heterogeneity of gliomas, with distinct regions such as active tumor, cystic and necrotic structures, and edema/invasion make the precise localization of tumor and its sub-regions a complex task. Manual segmentation of brain tumor is time-consuming and resource-intensive. Moreover, the segmentation results are prone to inter/intraobserver errors \cite{pereira2016brain}. Automated and accurate methods for precise localization of gliomas
% and its sub-regions from MRI scans are thus critical to diagnosis and therapy \cite{weller2021eano}.

Glioblastomas (GBM) are the most prevalent and aggressive primary brain tumors in adults. The significant morphological and histological diversity of gliomas, encompassing distinct zones like active tumors, cystic and necrotic structures, as well as edema and invasion areas, adds complexity to the precise identification of the tumor and its sub-regions. The manual segmentation of brain tumors demands substantial time and resources, while also being susceptible to errors originating from both inter and intra-observer variability \cite{pereira2016brain}. Consequently, developing automated and accurate methods to locate gliomas and their sub-regions within MRI scans precisely is paramount for diagnosis and treatment purposes \cite{weller2021eano}.

%Starting in 2012, the BraTS competition has established itself as a standard reference for brain MRI segmentation. 
The BraTS datasets and challenges offer an extensive repository of labeled brain MR images in an open-source format, facilitating the development of cutting-edge solutions in the field of neuroradiology. The BraTS-2023 challenge includes 4 mpMRI scans from each patient, including native (T1N), post-contrast T1-weighted (T1C), T2-weighted (T2W), and T2
Fluid Attenuated Inversion Recovery (T2F) volumes. The label mask comprised three categories: Edema (ED), Enhancing Tumor (ET), and Necrosis (NE). The Tumor Core (TC) is defined as the combined value of ET and NE, while the sum of ED, ET, and NE forms the Whole Tumor (WT).
% \textcolor{red}{ T1 (T1N), T2 (T2W), Fluid-Attenuated Inversion Recovery (FLAIR, T2F), and T1 Contrast-Enhanced 
% (T1CE, T1C). }

Prior BraTS challenges focused on adult brain diffuse astrocytoma \cite{menze2014multimodal,bakas2017advancing,bakas2018identifying,baid2021rsna}. The current edition of BraTS presents a series of challenges encompassing a range of tumor types, incomplete data, and technological factors. The focus of this paper will be mainly on two challenges, namely (a) Brain Metastases Dataset and, (b) the BraTS-Africa Dataset, where we achieved good performance in the validation phase of the BraTS 2023 Challenge.
% ADD A LINE WHY THESE TWO - IS THERE A SIMILARITY OR OVERLAP? 

\begin{figure}[t]
            \centering
              % include first image
              \includegraphics[width=1\linewidth]{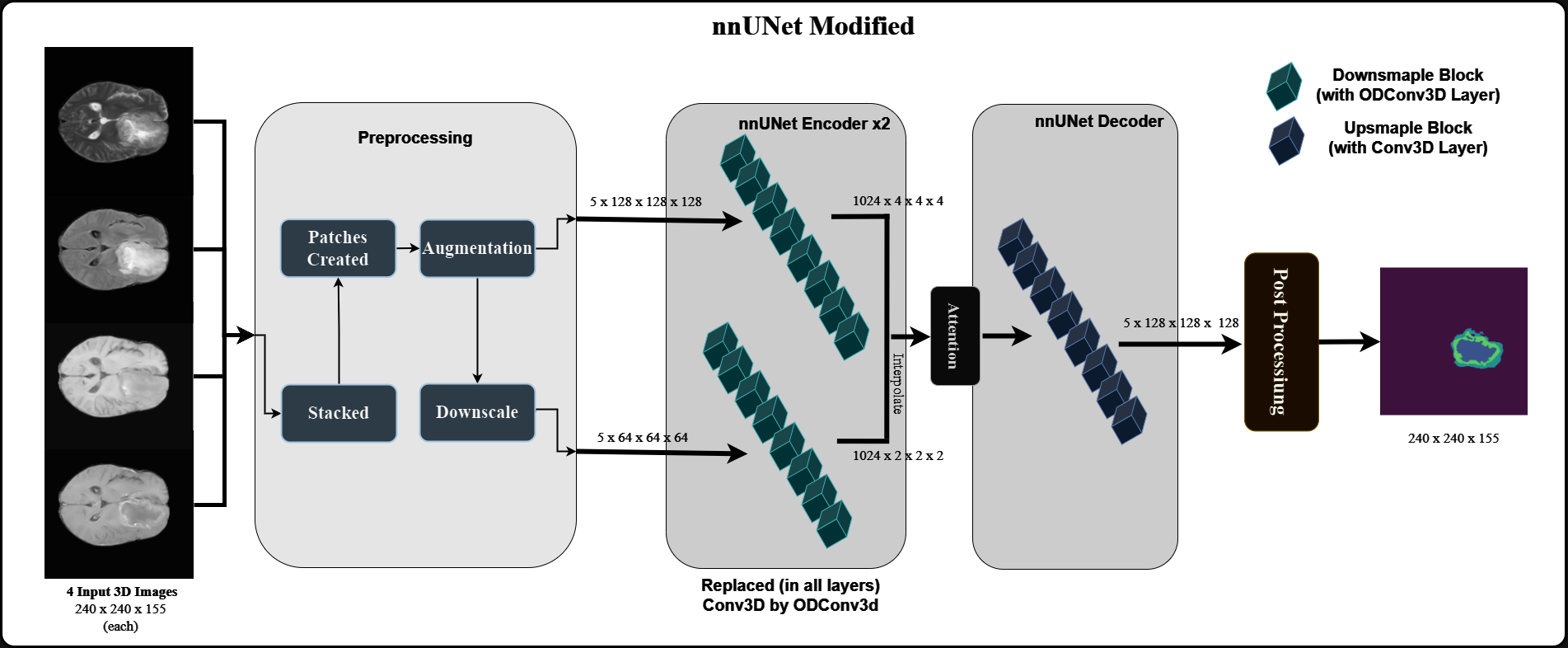}  
              \caption{\textbf{nnU-Net with the addition of Multi-Scale and ODConv3D Layers:} The architecture comprises a pair of identical encoders, responsible for extracting features from the input at distinct scales. These feature vectors undergo cross-attention processing, and the output is subsequently channeled into a decoder. The decoder's role involves upscaling the features to their original dimensions, effectively generating segmentations.}
            \label{architecture}
        \end{figure}

Our proposed approach uses the modified nnU-Net \cite{nnunet} architecture and comprises (a) two encoders, both inspired by the nnU-Net \cite{nnunet} framework (See Fig. \ref{architecture}) and (b) Omni-dimensional Dynamic Convolution (ODConv-2D) \cite{odconv} layers adapted to work with 3D images, which we refer as ODConv3D. 

By integrating two encoders into the network architecture, we exploit the capability of simultaneously processing the same image at two distinct scales. This approach enables us to extract features spanning a wide range of complexities.
% ADD A LINE ON WHY 2 ENCODERS.
ODConv utilizes an innovative multi-dimensional attention mechanism to acquire four distinct forms of attention for convolutional kernels, simultaneously encompassing all four dimensions of the kernel space. This enriched, dynamic convolution design enables the model to effectively capture intricate spatial, channel-wise, and temporal information. 
The results demonstrate the superiority of ODConv3D over the conventional convolutional layers in nnU-Net \cite{nnunet}. For submitting the code for evaluation purposes, we use MedPerf's MLCube\cite{karargyris2021medperf}.

\iffalse

  (You can add if you want more information about the general scope from the design document). Specifically, in the challenge described here, we focus on XYZ.

    Main points are(by Sahaj):
    \begin{itemize}
        \item We have used the nnU-Net architecture for the whole task and modified it in different ways.
        \item We duplicated the encoder network and trained one network with a normal image and the other with a downscaled image. At bottleneck, we applied cross attention between images and passed the image to the decoder whose architecture is the same as that of nnU-Net's decoder.
        \item We replaced the Conv3d of encoders by ODConv3D
        \item ODConv3D: an extension of ODConv2D(omni directional dynamic convolution, ICLR 2022) created by us.
        \item To create ODConv3D, we replaced the 2D layers with 3D layers and did some other required changes to make the layer work for voxels.
    \end{itemize}
    
    Clinical and technical motivation.
    
    Previous BraTS challenges References have been focused only on adult brain diffuse astrocytoma \cite{menze2014multimodal,bakas2017advancing,bakas2018identifying,baid2021rsna}. The focus of this year's BraTS cluster of challenges spans across various tumor entities, missing data, and technical considerations. (You can add if you want more information about the general scope from the design document). Specifically, in the challenge described here we focus on XYZ.

    Literature notes

    In this study ...

\fi

    \section{Materials}
        \subsection{Datasets}

    BraTS 2023 introduces five diverse challenges, each encompassing a distinct aspect of Brain Tumor Segmentation. Each challenge corresponds to datasets targeting specific types of brain tumors, namely: Adult Glioma, BraTS-Africa (Glioma from Sub-Saharan Africa), Meningioma, Brain Metastases, and Pediatric Tumors. 

    The BraTS dataset comprises retrospective multi-parametric magnetic resonance imaging (mpMRI) scans of brain tumors gathered from diverse medical institutions. These scans were acquired using varied equipment and imaging protocols, leading to a broad spectrum of image quality that reflects the diverse clinical approaches across different institutions. Annotations outlining each tumor sub-region were meticulously reviewed and validated by expert neuroradiologists.

    Following the customary methodology for evaluating machine learning algorithms, the BraTS 2023 challenge follows a division of the dataset into training, validation, and testing sets. The training data includes the provided ground truth labels. Participants are given access to the validation data, which lacks associated ground truth labels. The testing data, kept confidential during the challenge, serves as an evaluation benchmark.

    \subsubsection{Adult Glioma}

        The dataset comprises a comprehensive collection of 5,880 MRI scans from a cohort of 1,470 patients diagnosed with brain diffuse glioma \cite{gli}. Among these scans, 1,251 have been designated for training purposes, while an additional 219 scans are allocated for validation. This dataset is consistent with the data meticulously compiled for the BraTS 2021 Challenge. This dataset focuses on glioblastoma (GBM), which stands out as the most prevalent and aggressive form of brain-originating cancer. With a grade classification of IV, GBMs display an extraordinary degree of heterogeneity in terms of their appearance, morphology, and histology. Typically originating in the cerebral white matter, these tumors exhibit rapid growth and can attain considerable sizes before manifesting noticeable symptoms. Given the challenging nature of GBM and its dire prognosis, with a median survival rate of approximately 15 months, the comprehensive dataset is a vital resource for advancing research, diagnosis, and treatment strategies in the field.

    \subsubsection{BraTS-Africa}

        With a total training cohort size of 60 cases, the Sub-Saharan Africa dataset \cite{afr} represents a specialized collection of glioma cases among patients from the Sub-Saharan Africa region. This dataset stands out due to its distinct characteristics, arising from the utilization of lower-grade MRI technology and the limited availability of MRI scanners in the region. Consequently, the MRI scans included exhibit diminished image contrast and resolution. This reduction in clarity poses a challenge by obscuring the distinct features present in the aforementioned dataset and introducing intricacies in the segmentation process. Notably, patients within this dataset from the Sub-Saharan Africa region often present with comorbidities like HIV/AIDS, malaria, and malnutrition. These underlying health conditions influence the appearance of brain tumors in MRI scans, further complicating the diagnostic process.

    \subsubsection{Meningioma}

        The dataset contains an extensive collection of cases, comprising 1,000 instances designated for training and an extra 141 cases set aside for validation. These cases are centered around meningiomas \cite{men}, a specific type of tumor that originates from the meninges. These meninges are protective layers that cover the brain and spinal cord. Meningiomas are noteworthy since they represent the most frequently occurring primary tumors within the brain of adults. Their significant risks emphasize the clinical importance of these tumors in terms of health complications and mortality. This dataset was introduced as part of the BraTS challenge in 2023.

    \subsubsection{Brain Metastases}

        The Brain Metastases dataset's \cite{met} training segment comprises 165 samples. In comparison, an additional 31 samples are included in the validation set, providing a comprehensive collection of multiparametric MRI (mpMRI) scans focusing on brain tumors. Diverging from the characteristics of gliomas, which tend to be more readily detectable in their initial scans due to their larger size, brain metastases exhibit a distinctive heterogeneity in their dimensions. These metastatic tumors manifest across a spectrum of sizes, often presenting as smaller lesions within the brain. Notably, these metastatic growths possess the capacity to emerge at various locations throughout the brain, contributing to the diversity of their appearances.

        A noteworthy aspect of this dataset is the prevalence of brain metastases measuring less than 10 mm in diameter. These smaller lesions stand out in terms of their frequency, underscoring their clinical significance. Unlike their larger counterparts, these diminutive metastases have been observed to surpass others in terms of occurrence.

    \subsubsection{Pediatric Tumor}

        Despite certain similarities between pediatric tumors \cite{ped} and adult tumors, notable differences exist in their appearance in both imaging and clinical contexts. Both high-grade gliomas, such as GBMs, and pediatric diffuse midline gliomas (DMGs) have limited survival rates. Interestingly, DMGs are roughly three times rarer than GBMs. While GBMs typically emerge in the frontal or temporal lobes in the age of 60s, DMGs are predominantly found within pons and are frequently identified in the age range of 5 and 10.

        GBMs are typically characterized by the presence of an enhancing tumor region in post-gadolinium T1-weighted MRI scans, accompanied by necrotic regions. In contrast, these imaging characteristics are less pronounced or distinct in DMGs. Pediatric brain tumors are more inclined to be low-grade gliomas, generally displaying slower growth rates compared to those in adults. The dataset at hand comprises 99 training samples and 45 validation samples.

    \begin{figure}[t]
          \centering
          % include first image
          \includegraphics[width=1\linewidth]{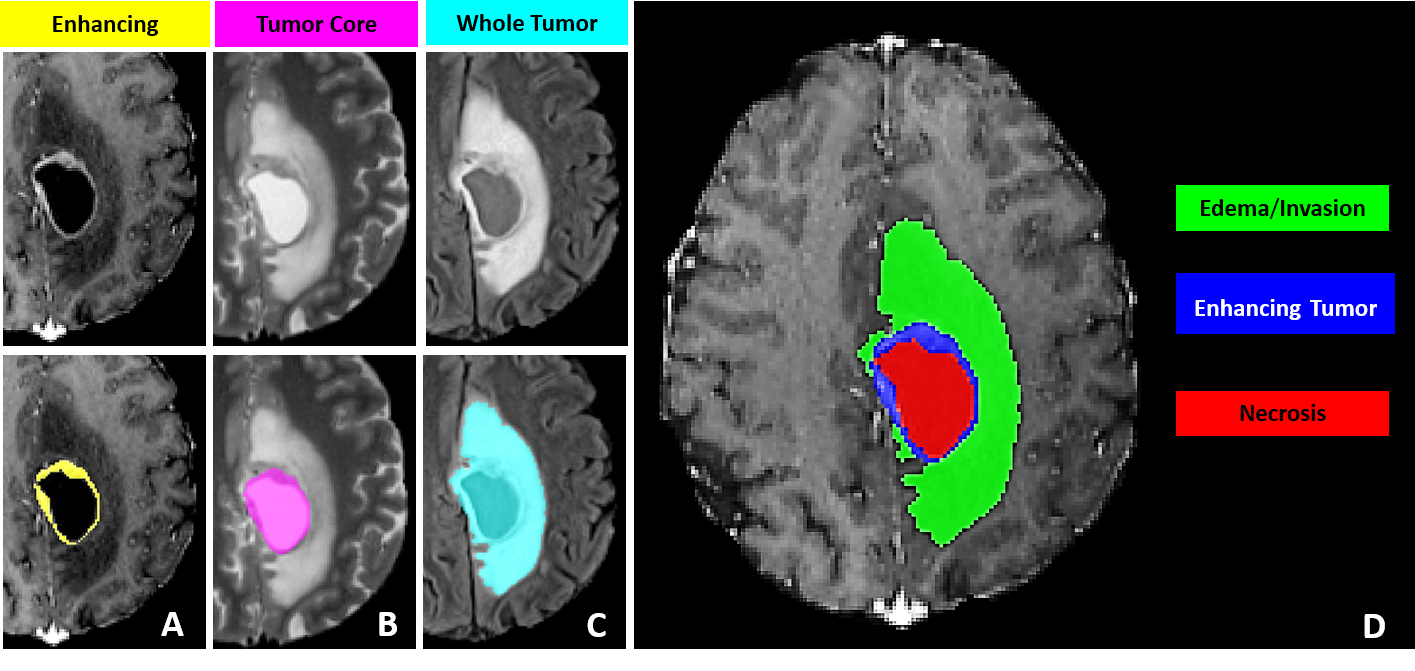}  
          \caption{\textbf{Tumor Sub-regions Annotations} Depicted various annotated tumor sub-regions across different multi-parametric MRI scans. Image panels A-C illustrate the areas designated for assessing algorithm performance, highlighting the enhancing tumor (ET - indicated in yellow) as seen in a T1C scan. This region encompasses the cystic/necrotic components of the core in panel A. In panels B and C, the tumor core (TC - magenta) and the entire tumor (WT - cyan) are delineated, respectively, in corresponding T2W and T2F scans. Panel D provides an overview of the merged segmentations forming the comprehensive tumor sub-region labels given to participants in the BraTS 2021 challenge. These labels comprise the enhancing core (yellow), necrotic/cystic core (red), and edema/invasion (green).
         The image is obtained from \cite{baid2021rsna}.}
        \label{annotations}
    \end{figure}
    
    \subsection{Annotation Methtod}       

        The dataset is established via a meticulous process for annotating tumor sub-regions, combining automated segmentation and manual refinement by expert neuroradiologists. The methodology involves initial automated segmentations using established methodsDeepMedic \cite{kamnitsas2017efficient, mckinley2018ensembles, isensee2020nnu}, which are then fused through the STAPLE label fusion \cite{warfield2004simultaneous} technique to address errors. Expert annotators meticulously enhance these segmentations using multimodal MRI scans and ITK-SNAP \cite{itksnap} software. Senior neuroradiologists review the refined segmentations, ensuring accuracy. This iterative process results in annotations conforming to BraTS-defined sub-regions: enhancing tumor (ET), tumor core (TC), and whole tumor (WT) as described in Fig. \ref{annotations}. The process acknowledges challenges in radiological tumor boundary definition, providing a standardized, dependable ground truth for assessing segmentation algorithms.

\section{Proposed Algorithm}
   In this section, we introduce our innovative methodology designed to precisely segment brain tumors using a multi-modal attention-based model. The proposed algorithm leverages a combination of data preprocessing, model architecture, and attention mechanisms to enhance the segmentation accuracy on 3D brain scans.
    
    \subsection{Preprocessing}

        \subsubsection{Data Augmentation}
            Before inputting the data into our model, we apply multiple data augmentation techniques to enhance the robustness of the model. These random augmentations within a specific range include crop, zoom, flip, noise, blur, brightness, and contrast. This ensures that the model can generalize better to various spatial transformations and variations in the input data.

        \subsubsection{Patch Extraction}
            We extract patches of size $5$ x $128$ x $128$ x $128$ from the original brain scan images. These patches capture localized features and enable the model to focus on specific regions of interest. This patch-based approach also aids in reducing memory requirements and computational complexity during training and inference.

    \subsection{Model Architecture}

        \subsubsection{Encoder}

            Our proposed model architecture consists of two encoder layers, each inspired by the nnU-Net \cite{nnunet} framework. Instead of using conventional Conv3D layers, we employ Omni-dimensional Dynamic Convolution 2D (ODConv2D) layers, which we extend for use with 3D images and refer to as ODConv3D. ODConv3D integrates a multi-dimensional attention mechanism utilizing a parallel strategy. This empowers us to acquire complementary attention patterns for convolutional kernels across all four dimensions of the kernel space. This enriched dynamic convolution design enables the model to effectively capture intricate spatial, channel-wise, and temporal information.

        \subsubsection{Multi-Modal Attention}

            Our multi-modal attention strategy enhances the model's ability to leverage complementary information from different scales. We provide the original input image and a downsampled version of the same image as inputs to the model. This enables the model to capture fine-grained details and broader contextual information simultaneously. Separate encoders process each of these inputs, and the attention mechanism is then applied to fuse the encoded features. This approach encourages the model to focus on informative regions across scales, contributing to enhanced feature representation.

        \subsubsection{Decoder}

            The decoder of our model is similar to the nnU-Net's decoder architecture. It reconstructs the fused encoded features to generate a comprehensive feature map that preserves spatial information. This map is then processed to yield the final predicted segmentation.

    \subsection{Segmentation Process}

        \subsubsection{Post-Processing}

            Post-processing is applied to the output feature map to refine the predicted segmentation. This includes steps such as thresholding, morphological operations, and connected component analysis to remove noise and ensure coherent tumor regions.

        \subsubsection{Performance Metrics}

            To evaluate the accuracy and efficacy of our proposed model, we employ two performance metrics i.e., Dice coefficient and Hausdorff distance. These metrics provide quantitative measures of the segmentation quality and the model's ability to delineate tumor boundaries accurately.

    \subsection{Experimental Setup}

        We validate the performance of our proposed algorithm on the Brain Metastases dataset of brain scans containing instances of brain tumors. Each dataset instance consists of four 3D images of size $240$ x $240$ x $155$, along with corresponding ground truth segmentations. We divide the dataset into training and validation sets, and test on the synapse page, ensuring an unbiased evaluation of the model's performance.
        
    % The dataset contains instances as four 3D images(scans) of the brain of size $240$ x $240$ x $155$ along with a segmentation ground truth file of the same size. These images are stacked together and a patch of size $5$ x $128$ x $128$ x $128$ is created. Multiple augmentations are applied to the image before passing it to the model. 
    
    % The model consists of two encoder layers each similar to that of nnUNet with the replacement of Conv3D layer by ODConv3D layers. Each layer takes in separate images as input. Attention is applied to these separately encoded images and the resultant image is passed to the decoder. The decoder of the model is same as that of nnUNet. The decoded images is the processed to get the predicted segmentation.

    \section{Results}
            In this section, we present the results of our proposed %multi-modal 
    attention-based tumor segmentation model on two distinct datasets: the Brain Metastases Dataset and the BraTS-Africa Dataset. We compare our model's performance against several baselines, including the original nnU-Net architecture, variations incorporating ODConv3D and multiscale strategies, and a data processing combination approach.

    \subsection{Evaluation Metrics}

        We evaluate the segmentation accuracy using a set of standard performance metrics, including the Dice coefficient for overall segmentation, as well as specific metrics for lesion subregions, namely Lesion ET (Enhancing Tumor), Lesion TC (Tumor Core), and Lesion WT (Whole Tumor).

    %%%%%%%%%%%%%%%%%%%%%%%%%%%%%%%%%%%%%%%%%%%%%%%%%%%%%%%%%%%%%%%%%%%%%%%%%%%%%%%%%%%%%%%%%%%%%%%%%%%%%%%

    \subsection{BraTS-Africa Dataset}

        \subsubsection{Baseline Models}

            We evaluate the baseline performance of the nnU-Net architecture on the BraTS-Africa Dataset. As shown in Table 1, the nnU-Net achieves a Dice coefficient of 0.8980 for overall segmentation. Lesion ET, TC, and WT achieve Dice coefficients of 0.8354, 0.8485, and 0.6578, respectively.

        \subsubsection{Multiscale and ODConv3D Integration}

            We investigate the effects of integrating the Multiscale and ODConv3D into the nnU-Net architecture on the BraTS-Africa Dataset. The nnU-Net + Multiscale + ODConv3D model achieves an enhanced Dice coefficient of 0.9092 for overall segmentation. However, individual lesion subregions show varying outcomes, with Lesion ET experiencing a slightly decreased Dice coefficient.

        \subsubsection{Additional Post Processing}

            When we replace the labels of the nnU-Net model by the nnU-Net + Multiscale + ODConv3D model's Label 2 for the BraTS-Africa Dataset, this leads to an increase in Lesion WT from 0.65 to 0.75 as shown in table \ref{afr}, improving the overall accuracy. Our model's performance is highlighted in Table \ref{afr_com}, demonstrating competitive results. Across all lesion subtypes, our model achieves Lesion ET scores of 0.8354, Lesion TC scores of 0.8485, and Lesion WT scores of 0.7564.
%%%%%%%%%%%%%%%%%%%%%%%%%%%%%%%%%%%%%%%%%%%%%%%%%%%%%%%%%%%%%%%%%%%%%%%%%%%%%%%%%%%%%%%%%%%%%%%%%%%%%%%%%%%%%%%%%

\begin{table}[!t]
\centering
\caption{Results of the different algorithms on  BraTS-Africa validation dataset. Here a + b is taking best of a and b.}
\small
\begin{tabular}{|p{2.2in}|p{0.95in}|p{0.95in}|p{0.95in}|p{0.95in}|}  \hline
\textbf{Algorithm} &\textbf{Training Dice*} & \textbf{Lesion ET} & \textbf{Lesion TC} & \textbf{Lesion WT}\\ \hline \hline
\textbf{nnU-Net (a)}                       & 0.8980 & 0.8354 & 0.8485 & 0.6578 \\ \hline
\textbf{nnU-Net + Multiscale + ODConv3D (b)} & 0.9092 & 0.8082 & 0.7634 & \textbf{0.7872} \\ \hline
\textbf{a + b}                           &        & \textbf{0.8354} & \textbf{0.8485} & 0.7564 \\ \hline           
\end{tabular}
\footnotesize{ * The models are trained on Glioma Dataset}
\label{afr}
\end{table}

%%%%%%%%%%%%%%%%%%%%%%%%%%%%%%%%%%%%%%%%%%%%%%%%%%%%%%%%%%%%%%%%%%%%%%%%%%%%%%%%%%%%%%%%%%%%%%%%%%%%%%%%%%%%%%

\begin{table}[!t]
\centering
\caption{Results comparison of the different teams on BraTS-Africa validation dataset.}
\small
\begin{tabular}{|p{2.2in}|p{0.95in}|p{0.95in}|p{0.95in}|}  \hline
\textbf{Algorithm} & \textbf{Lesion ET} & \textbf{Lesion TC} & \textbf{Lesion WT}\\ \hline \hline
{SPARC}  & 0.7478 & 0.7649 & 0.7515 \\ \hline
{@harshi}  & 0.7537 & 0.7051 & 0.5991 \\ \hline
{@ntnu40940111s}  & 0.7603 & 0.7695 & 0.8403 \\ \hline
{blackbean}  & 0.8264 & 0.8464 & 0.5690 \\ \hline
{@nic-vicorob}  & 0.8029 & 0.7985 & 0.7487 \\ \hline
{BraTS2023\_SPARK\_UNN}  & 0.7577 & 0.7907 & \textbf{0.8647} \\ \hline
\textbf{Ours} & \textbf{0.8354} & \textbf{0.8485} & 0.7564 \\ \hline           
\end{tabular}
\label{afr_com}
\end{table}

%%%%%%%%%%%%%%%%%%%%%%%%%%%%%%%%%%%%%%%%%%%%%%%%%%%%%%%%%%%%%%%%%%%%%%%%%%%%%%%%%%%%%%%%%%%%%%%%%%%%%%%%%%%%%%%

    \subsection{Brain Metastases Dataset}

        \subsubsection{Baseline Models}

            We begin by evaluating the baseline performance of the nnU-Net architecture on the Brain Metastases Dataset. As depicted in Table 3, the nnU-Net achieves a Dice coefficient of 0.7675 for overall segmentation. The performance varies across lesion subregions, with Lesion ET, Lesion TC, and Lesion WT achieving Dice coefficients of 0.5157, 0.5105, and 0.4656, respectively.

        \subsubsection{Impact of ODConv3D}

            We assess the effectiveness of Omni-dimensional Dynamic Convolution (ODConv3D) by introducing the nnU-Net + ODConv3D model. This model replaces traditional Conv layers with ODConv3D layers for enhanced feature extraction. The results in Table 3 demonstrate that ODConv3D leads to an improved Dice coefficient of 0.7953 for overall segmentation. This improvement extends to lesion subregions, with Lesion ET, Lesion TC, and Lesion WT achieving Dice coefficients of 0.5364, 0.5451, and 0.5141, respectively.

        \subsubsection{Multiscale Strategy}

            We explore the benefit of a multiscale strategy by introducing the nnU-Net + Multiscale model. This approach leverages multiple scales of the input data to improve feature representation. As indicated in Table \ref{met}, the multiscale strategy yields a Dice coefficient of 0.7771 for overall segmentation. Lesion TC substantially improves with a Dice coefficient of 0.5737, compared to other subregions.

        \subsubsection{Combined Approach}

            The culmination of our model's innovation comes with the nnU-Net + Multiscale + ODConv3D configuration. Combining the multiscale strategy and ODConv3D layers yields exceptional results, evident in the overall Dice coefficient of 0.8188. This combined approach significantly enhances the segmentation performance across lesion subregions, with Dice coefficients of 0.5896, 0.6406, and 0.5555 for Lesion ET, Lesion TC, and Lesion WT, respectively. Table \ref{met_com} showcases our model's performance, underscoring its competitive results. Our model consistently achieves good scores across all lesion subtypes, including Lesion ET with a score of 0.5896, Lesion TC with a score of 0.6406, and Lesion WT with a score of 0.5648.

%%%%%%%%%%%%%%%%%%%%%%%%%%%%%%%%%%%%%%%%%%%%%%%%%%%%%%%%%%%%%%%%%%%%%%%%%%%%%%%%%%%%%%%%%%%%%%%%%%%%%%%%%

    \begin{table}[!t]
        \centering
        \caption{Results of the different algorithms on Brain Metastases validation dataset.}
        \small
        \begin{tabular}{|p{2.2in}|p{0.95in}|p{0.95in}|p{0.95in}|p{0.95in}|}  \hline
        \textbf{Algorithm} &\textbf{Training Dice} & \textbf{Lesion ET} & \textbf{Lesion TC} & \textbf{Lesion WT}\\ \hline \hline
        \textbf{nnU-Net}                       & 0.7675 & 0.5157 & 0.5105 & 0.4656 \\ \hline
        \textbf{nnU-Net + ODConv3D}              & 0.7953 & 0.5364 & 0.5451 & 0.5141 \\ \hline
        \textbf{nnU-Net + Multiscale}          & 0.7771 & 0.5354 & 0.5737 & 0.5008 \\ \hline
        \textbf{nnU-Net + Multiscale + ODConv3D} & \textbf{0.8188} & \textbf{0.5896} & \textbf{0.6406} & \textbf{0.5555} \\ \hline           
        \end{tabular}
        \label{met}
    \end{table}

%%%%%%%%%%%%%%%%%%%%%%%%%%%%%%%%%%%%%%%%%%%%%%%%%%%%%%%%%%%%%%%%%%%%%%%%%%%%%%%%%%%%%%%%%%%%%%%%%%%%%%%%%%%%%%%%

\begin{table}[!t]
\centering
\caption{Results comparison of the different teams on Brain Metastases validation dataset.}
\small
\begin{tabular}{|p{2.2in}|p{0.95in}|p{0.95in}|p{0.95in}|}  \hline
\textbf{Algorithm} & \textbf{Lesion ET} & {Lesion TC} & \textbf{Lesion WT}\\ \hline \hline
{SPARC}  & 0.4133 & 0.4378 & 0.4698 \\ \hline
{MIA\_SINTEF}  & 0.4433 & 0.4774 & 0.4832 \\ \hline
{DeepRadOnc}  & 0.4575 & 0.4913 & 0.4665 \\ \hline
{@jeffrudie}  & 0.4119 & 0.5171 & 0.5168 \\ \hline
{@parida12}  & 0.5592 & 0.6039 & \underline{0.5650} \\ \hline
{CNMC\_PMI2023}  & \textbf{0.608} & \textbf{0.649} & \textbf{0.587} \\ \hline
\textbf{Ours} & \underline{0.5896} & \underline{0.6406} & 0.5555 \\ \hline           
\end{tabular}
\label{met_com}
\end{table}

    \subsection{Comparison across Diverse Datasets}
        
        To comprehensively assess our proposed multi-modal attention-based tumor segmentation model, we conducted evaluations on a diverse range of challenges: Adult Glioma Segmentation, BraTS-Africa Segmentation, Meningioma Segmentation, Brain Metastases Segmentation, and Pediatric Tumors Segmentation. As summarized in table \ref{compare_training_challenge} and \ref{compare_testing_challenge}, the results offer a comparison of the achieved Dice coefficients across different datasets in the validation and testing phase of the BraTS 2023 challenge.
        
        Moreover, to provide visual clarity on the model's performance, we present figures in table \ref{compare}. These figures visually represent the segmentation outcomes on representative examples from each challenge. These visual insights enrich our understanding of how the model's capabilities translate into actual segmentations in various clinical scenarios.

%%%%%%%%%%%%%%%%%%%%%%%%%%%%%%%%%%%%%%%%%%%%%%%%%%%%%%%%%%%%%%%%%%%%%%%%%%%%%%%%%%%%%%%%%%%%%%%%%%%%%%%%%%%%%%

\begin{table}[!t]
\centering
\caption{Scores of modified nnU-Net on validation datasets in Brats 2023 Challenge.}
\small
\begin{tabular}{|p{2.2in}|p{0.95in}|p{0.95in}|p{0.95in}|p{0.95in}|}  \hline
\textbf{Challenge} &\textbf{Training Dice} & \textbf{Lesion ET} & \textbf{Lesion TC} & \textbf{Lesion WT}\\ \hline \hline
\textbf{Adult Glioma Segmentation} & 0.909 & 0.798 & 0.826 & 0.789 \\ \hline
\textbf{BraTS-Africa Segmentation} & 0.909 & 0.835 & 0.849 & 0.756 \\ \hline
\textbf{Meningioma Segmentation} & 0.889 & 0.783 & 0.780 & 0.756 \\ \hline
\textbf{Brain Metastases Segmentation} & 0.819 & 0.590 & 0.641 & 0.556 \\ \hline
\textbf{Pediatric Tumors Segmentation} & 0.805 & 0.565 & 0.764 & 0.813 \\ \hline
\end{tabular}
\label{compare_training_challenge}
\end{table}

%%%%%%%%%%%%%%%%%%%%%%%%%%%%%%%%%%%%%%%%%%%%%%%%%%%%%%%%%%%%%%%%%%%%%%%%%%%%%%%%%%%%%%%%%%%%%%%%%%%%%%%%%%%%%%%

\begin{table}[!t]
\centering
\caption{Scores of modified nnU-Net on testing datasets in Brats 2023 Challenge.}
\small
\begin{tabular}{|p{2.2in}|p{0.95in}|p{0.95in}|p{0.95in}|}  \hline
\textbf{Challenge} & \textbf{Lesion ET} & \textbf{Lesion TC} & \textbf{Lesion WT}\\ \hline \hline
\textbf{Adult Glioma Segmentation} & 0.798 & 0.826 & 0.789 \\ \hline
\textbf{BraTS-Africa Segmentation} & 0.818 & 0.775 & 0.845 \\ \hline
\textbf{Meningioma Segmentation} & 0.799 & 0.773 & 0.763 \\ \hline
\textbf{Brain Metastases Segmentation} & 0.491 & 0.534 & 0.483 \\ \hline
\textbf{Pediatric Tumors Segmentation} & 0.480 & 0.320 & 0.347 \\ \hline
\end{tabular}
\label{compare_testing_challenge}
\end{table}

%%%%%%%%%%%%%%%%%%%%%%%%%%%%%%%%%%%%%%%%%%%%%%%%%%%%%%%%%%%%%%%%%%%%%%%%%%%%%%%%%%%%%%%%%%%%%%%%%%%%%%%%%%%%%%%0

\begin{table}[!h]
    \centering
    \begin{tabular}{ccccccc}
         & \textbf{t2w} & \textbf{t2f} & \textbf{t1n} & \textbf{t1c} & \textbf{Ground Truth} & \textbf{Predictions} \\
        (a) &
        \includegraphics[width=2.5cm]{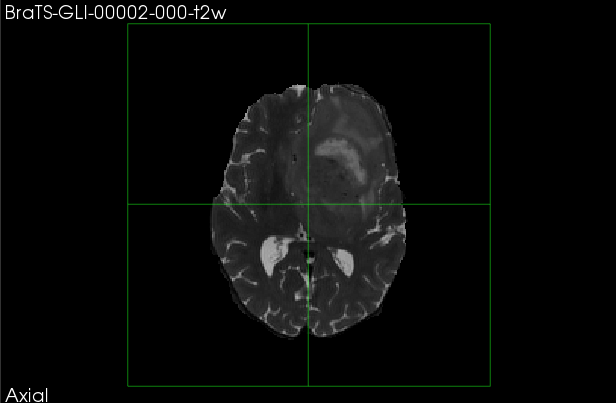} &
        \includegraphics[width=2.5cm]{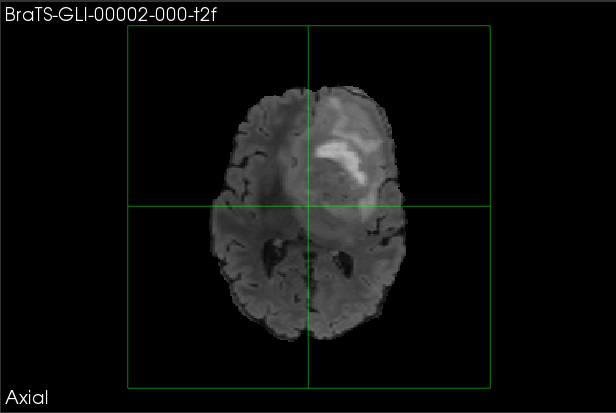} &
        \includegraphics[width=2.5cm]{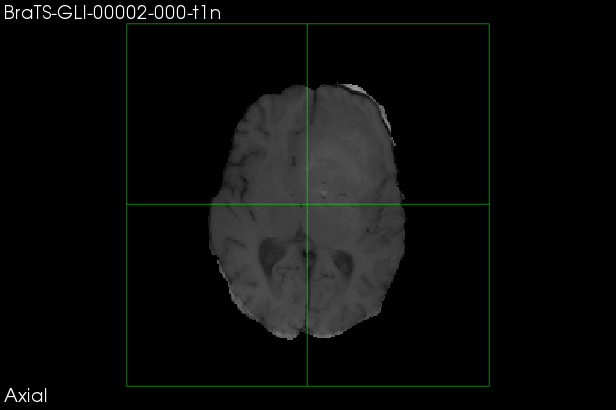} &
        \includegraphics[width=2.5cm]{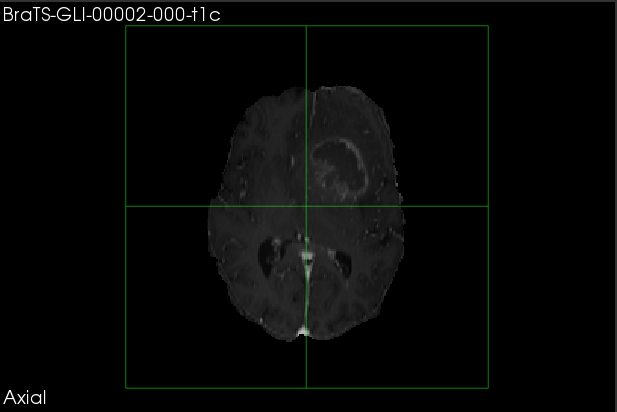} &
        \includegraphics[width=2.5cm]{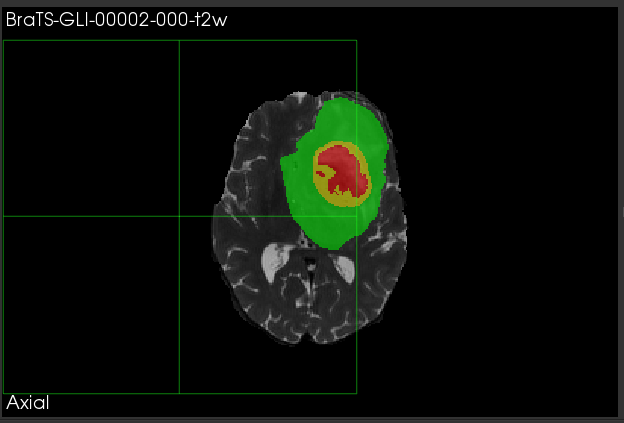}&
        \includegraphics[width=2.5cm]{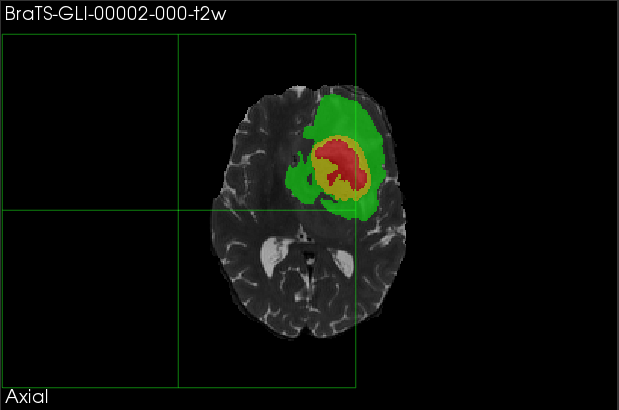} \\
    
        (b) &
        \includegraphics[width=2.5cm]{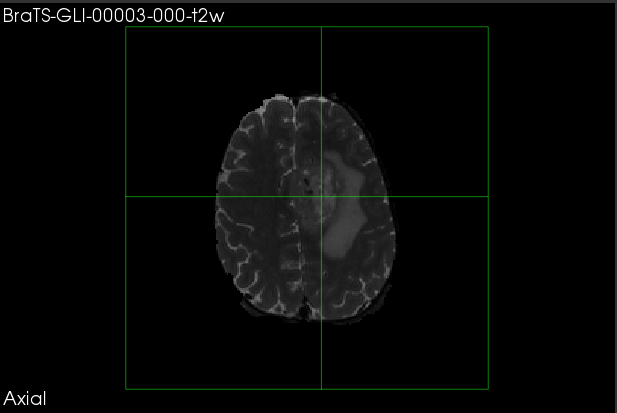} &
        \includegraphics[width=2.5cm]{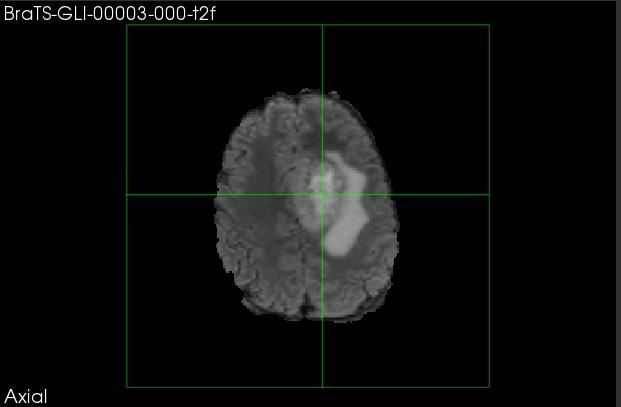} &
        \includegraphics[width=2.5cm]{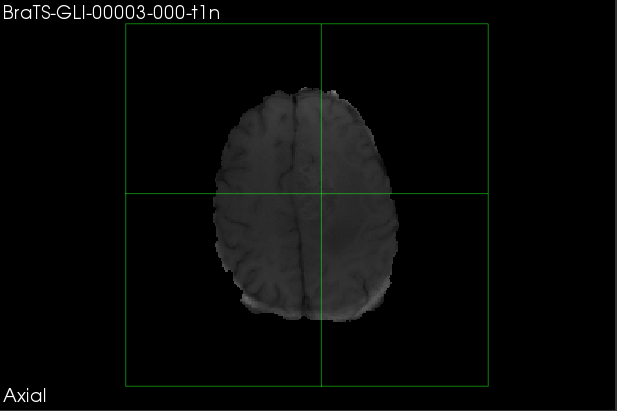} &
        \includegraphics[width=2.5cm]{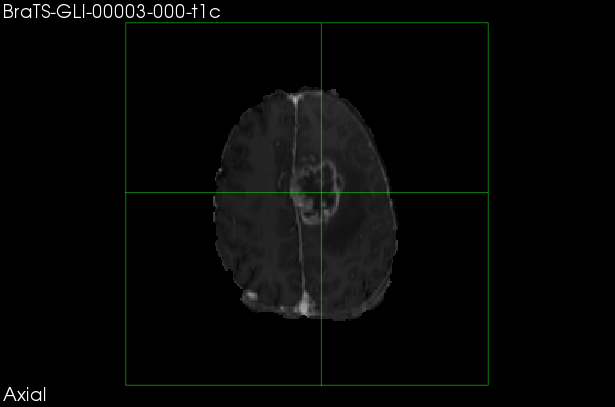}&
        \includegraphics[width=2.5cm]{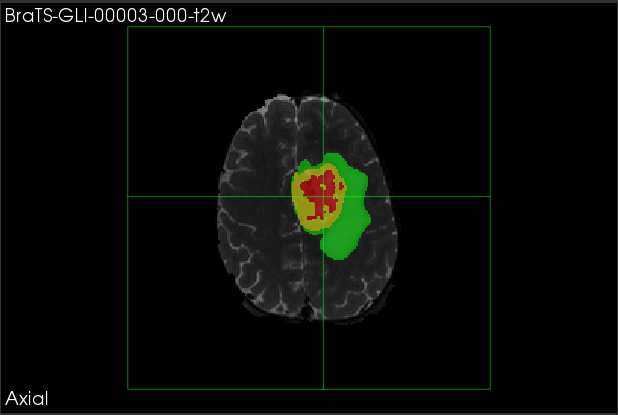}&
        \includegraphics[width=2.5cm]{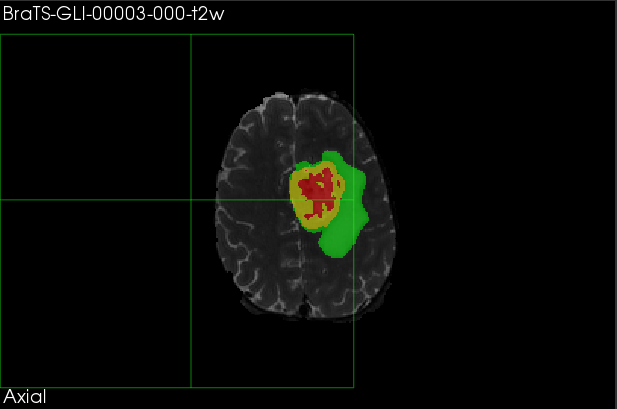} \\

        (c) &
        \includegraphics[width=2.5cm]{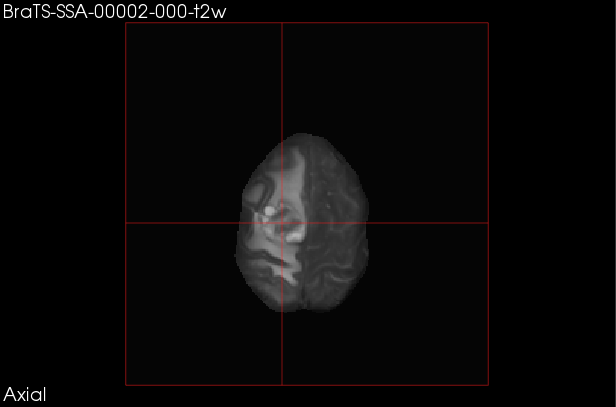} &
        \includegraphics[width=2.5cm]{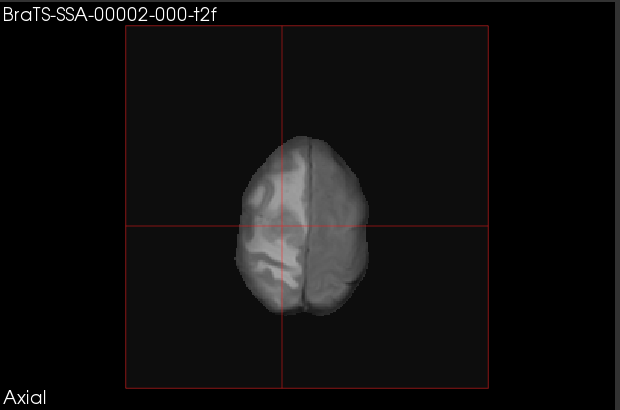} &
        \includegraphics[width=2.5cm]{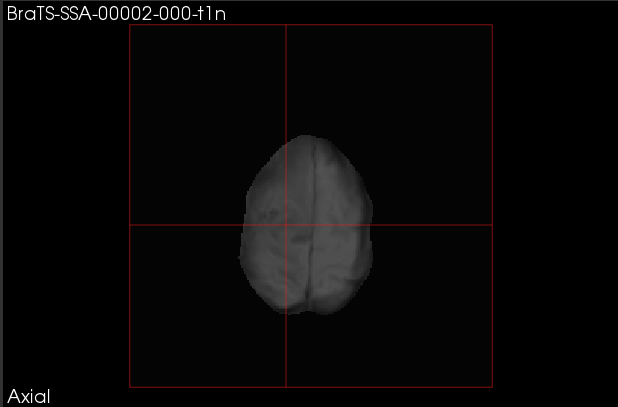} &
        \includegraphics[width=2.5cm]{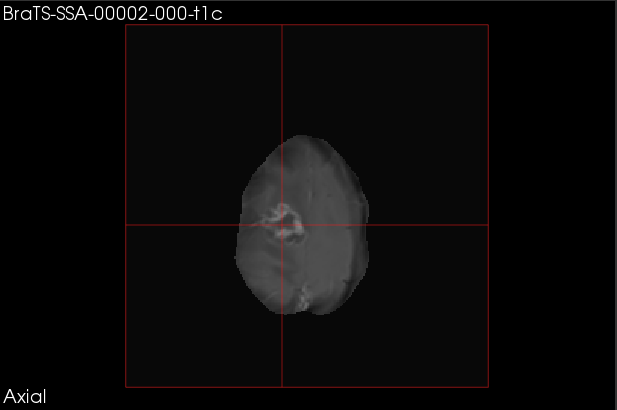}&
        \includegraphics[width=2.5cm]{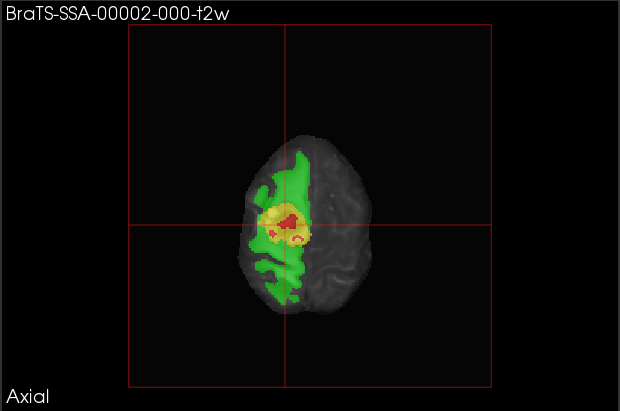} &
         \includegraphics[width=2.5cm]{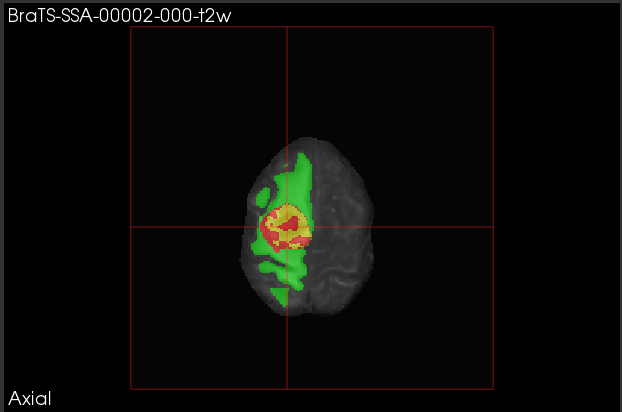}\\
    
        (d) &
        \includegraphics[width=2.5cm]{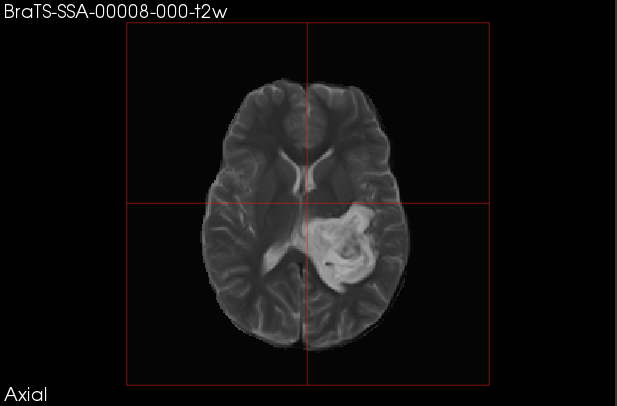} &
        \includegraphics[width=2.5cm]{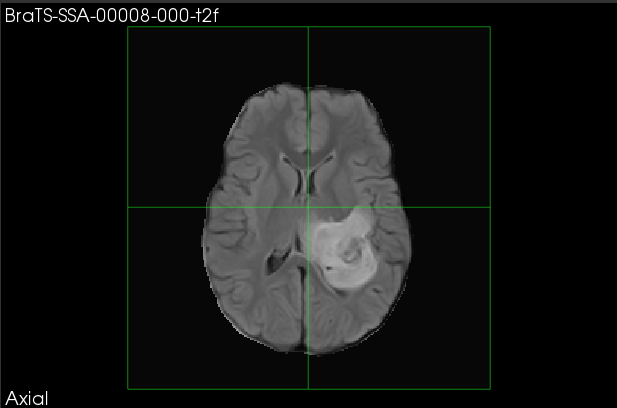} &
        \includegraphics[width=2.5cm]{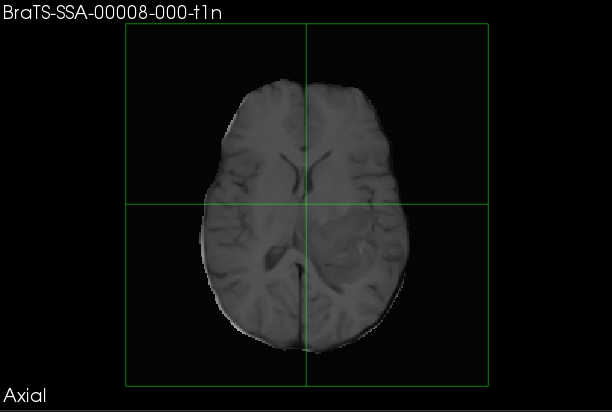} &
        \includegraphics[width=2.5cm]{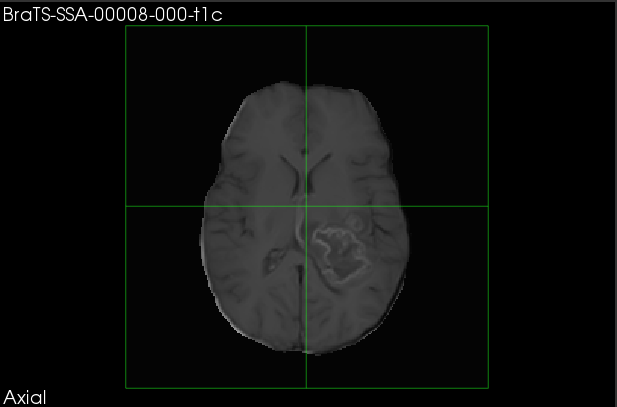}&
        \includegraphics[width=2.5cm]{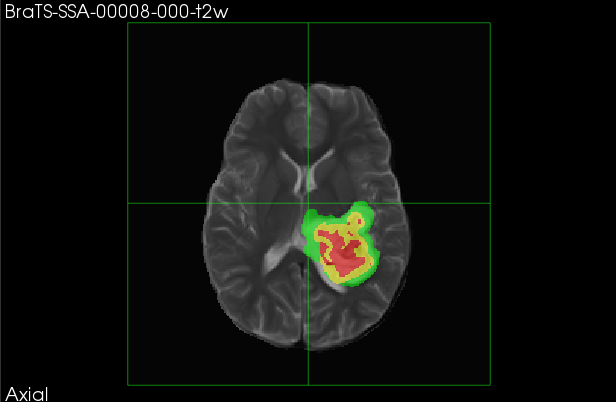}&
        \includegraphics[width=2.5cm]{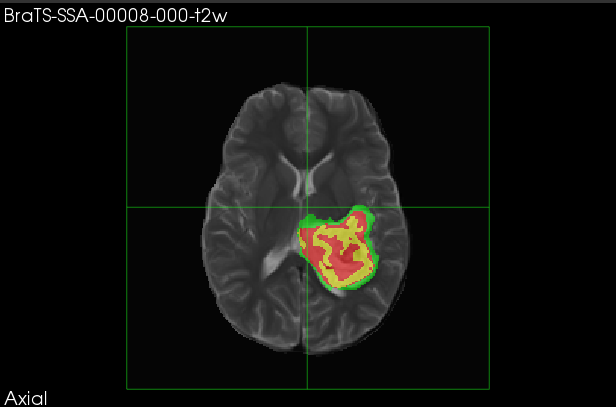} \\

        (e) &
        \includegraphics[width=2.5cm]{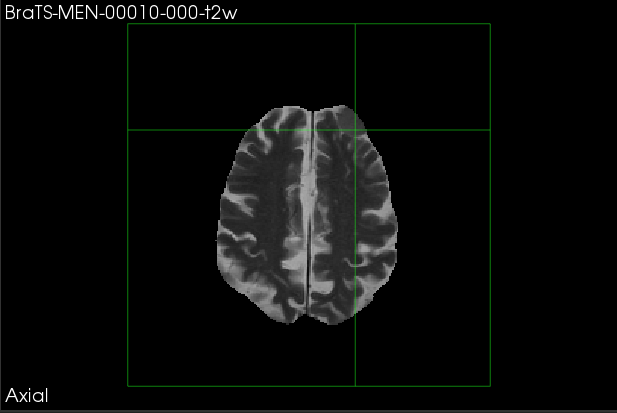} &
        \includegraphics[width=2.5cm]{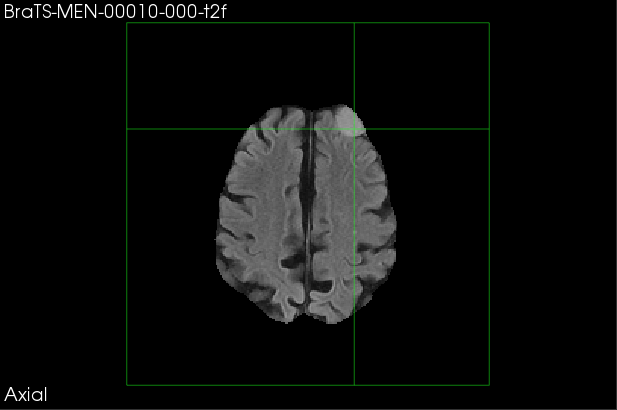} &
        \includegraphics[width=2.5cm]{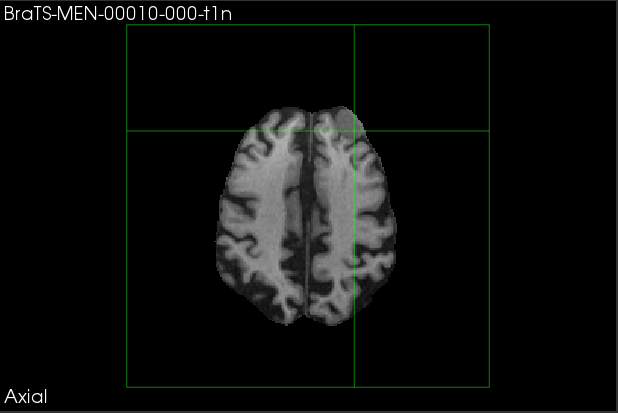} &
        \includegraphics[width=2.5cm]{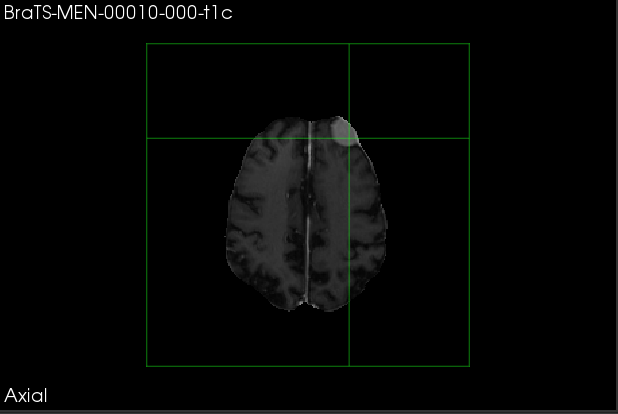}&
        \includegraphics[width=2.5cm]{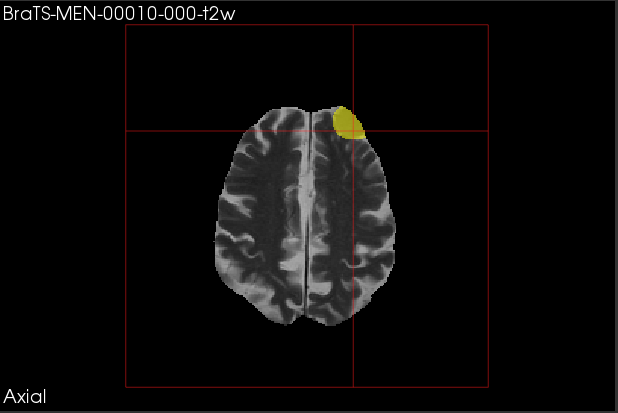}&
        \includegraphics[width=2.5cm]{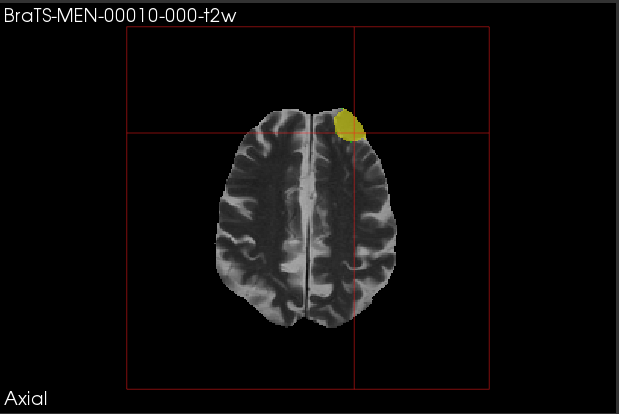} \\
    
        (f) &
        \includegraphics[width=2.5cm]{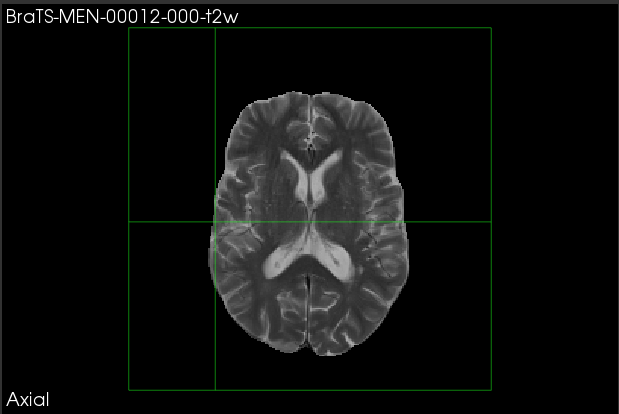}&
        \includegraphics[width=2.5cm]{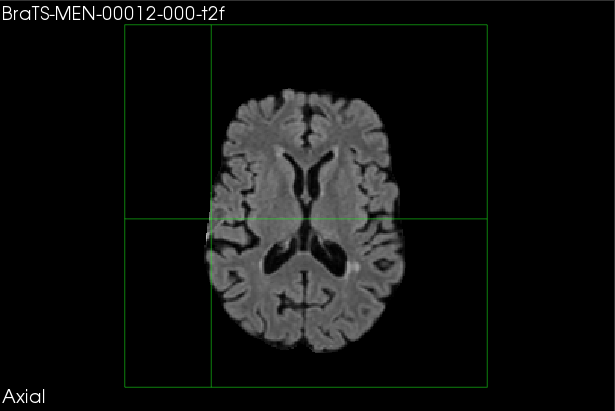} &
        \includegraphics[width=2.5cm]{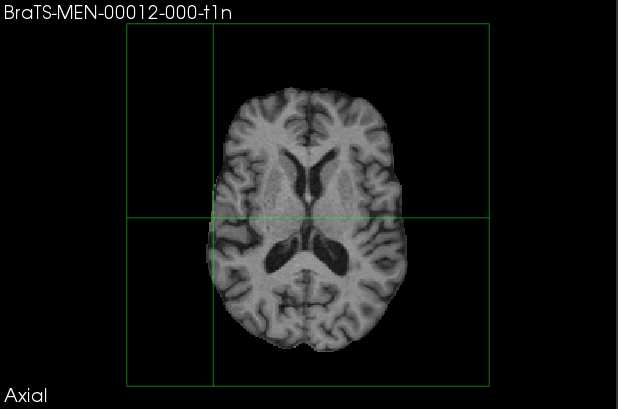} &
        \includegraphics[width=2.5cm]{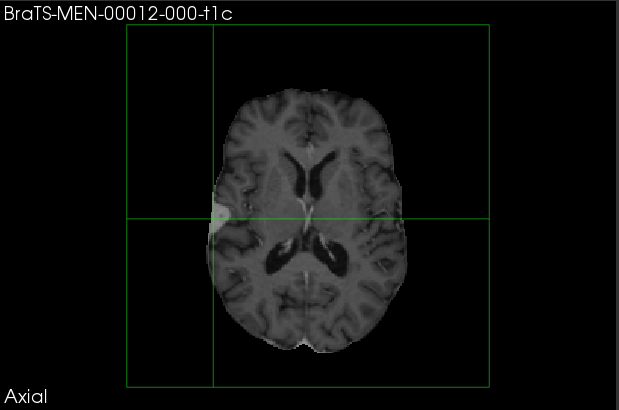} &
        \includegraphics[width=2.5cm]{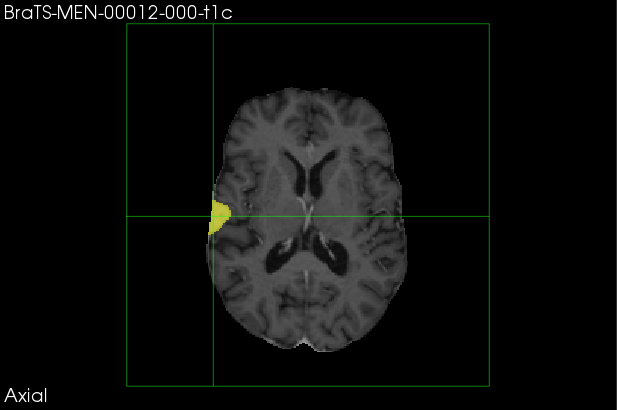}&
        \includegraphics[width=2.5cm]{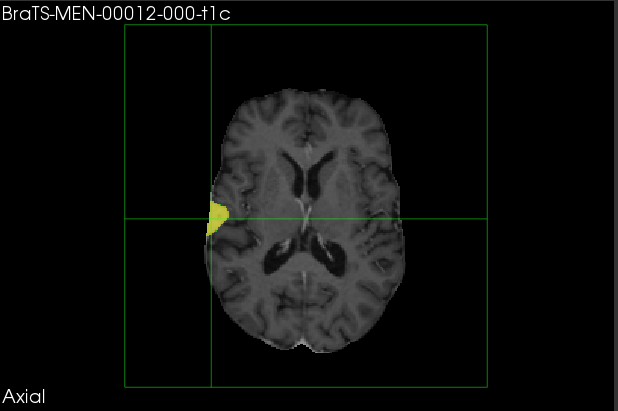} \\

        (g) &
        \includegraphics[width=2.5cm]{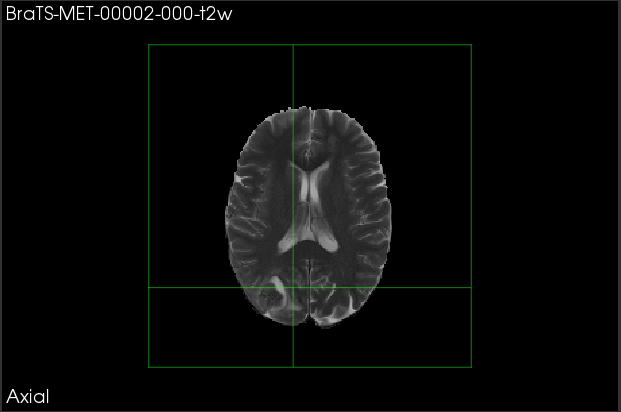} &
        \includegraphics[width=2.5cm]{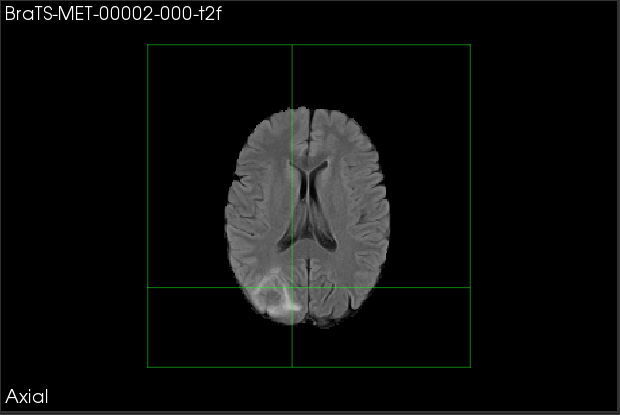} &
        \includegraphics[width=2.5cm]{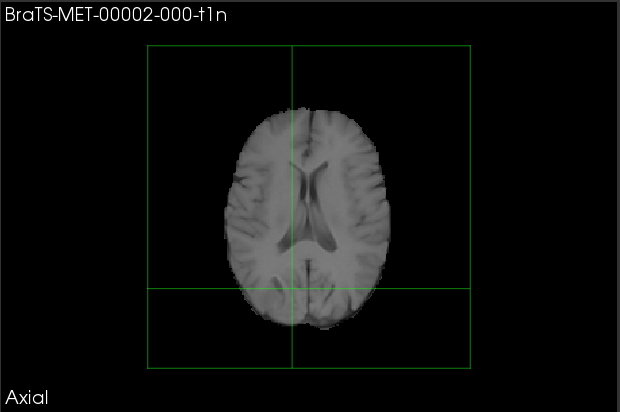} &
        \includegraphics[width=2.5cm]{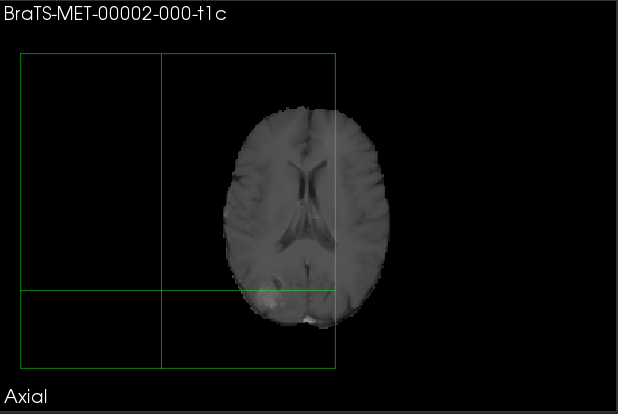} &
        \includegraphics[width=2.5cm]{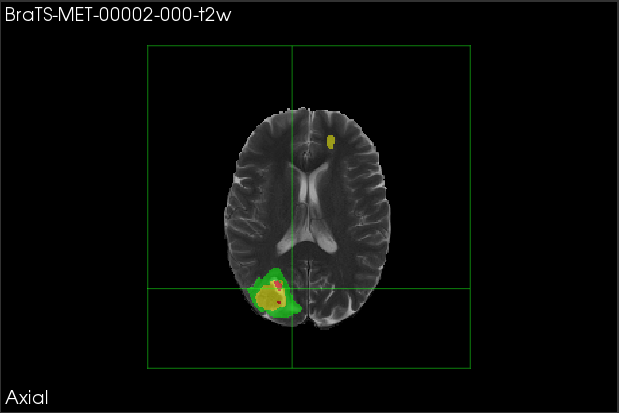} &
        \includegraphics[width=2.5cm]{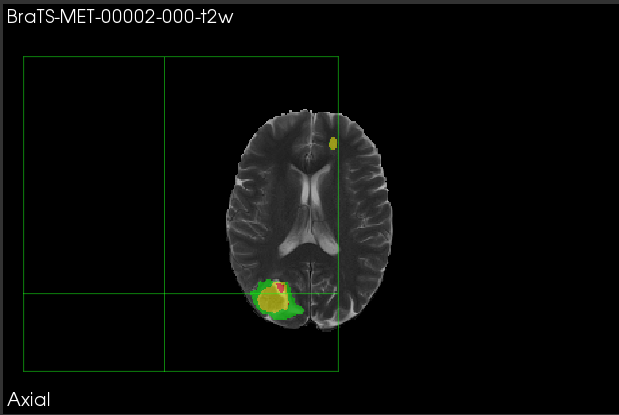} \\
    
        (h) &
        \includegraphics[width=2.5cm]{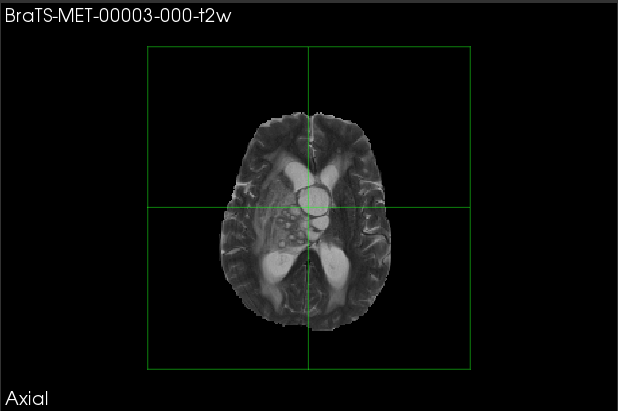} &
        \includegraphics[width=2.5cm]{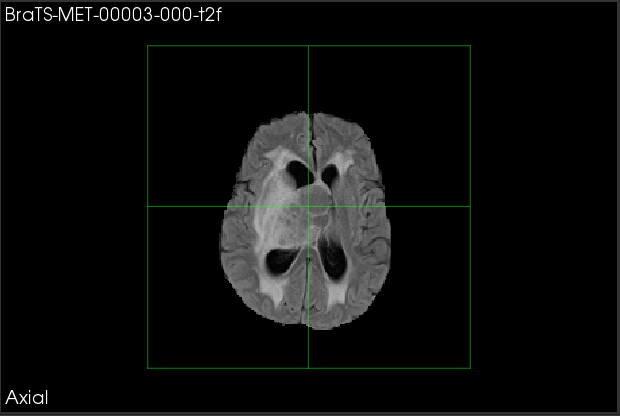} &
        \includegraphics[width=2.5cm]{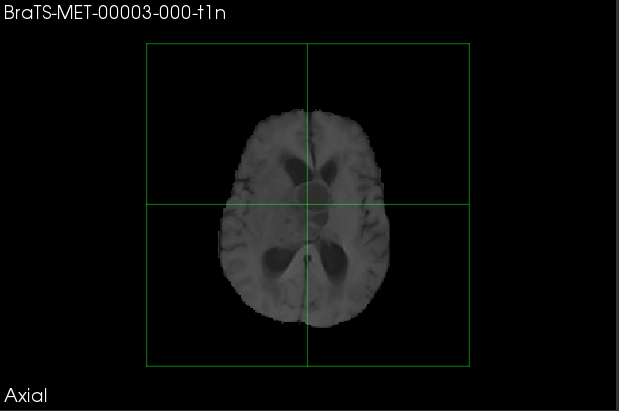} &
        \includegraphics[width=2.5cm]{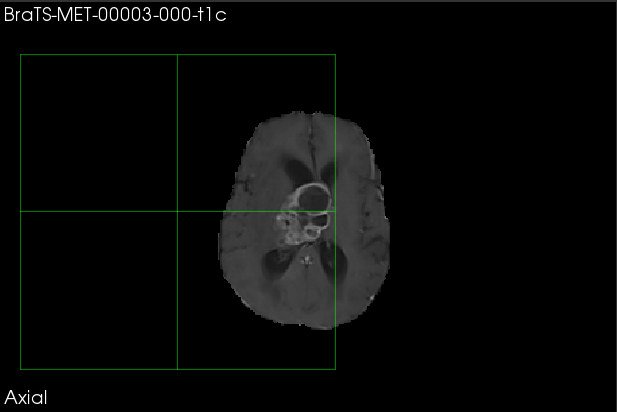}&
        \includegraphics[width=2.5cm]{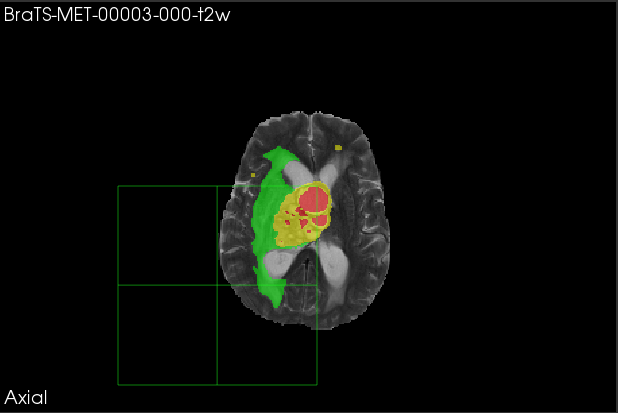}&
        \includegraphics[width=2.5cm]{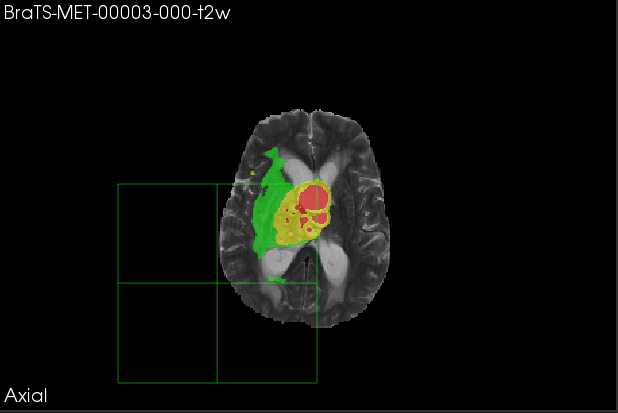} \\

        (i) &
        \includegraphics[width=2.5cm]{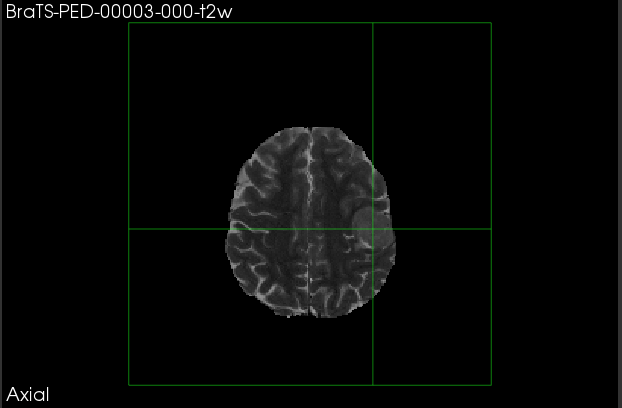} &
        \includegraphics[width=2.5cm]{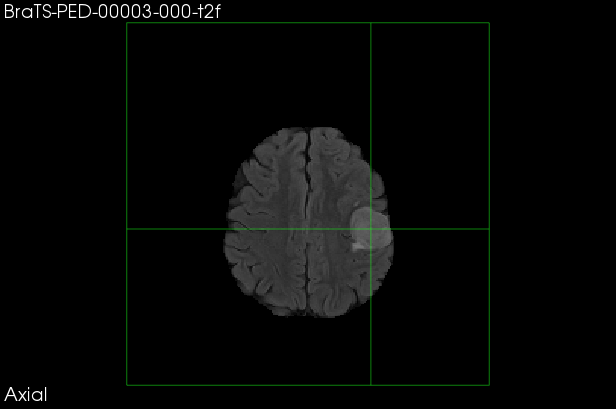} &
        \includegraphics[width=2.5cm]{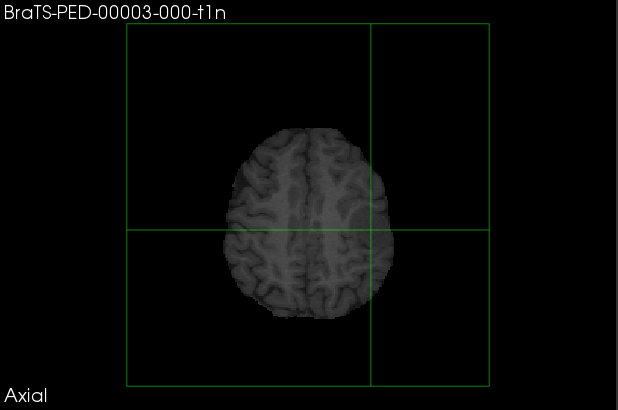} &
        \includegraphics[width=2.5cm]{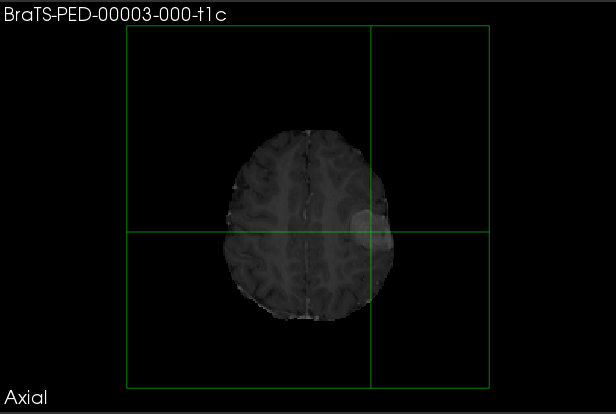}&
        \includegraphics[width=2.5cm]{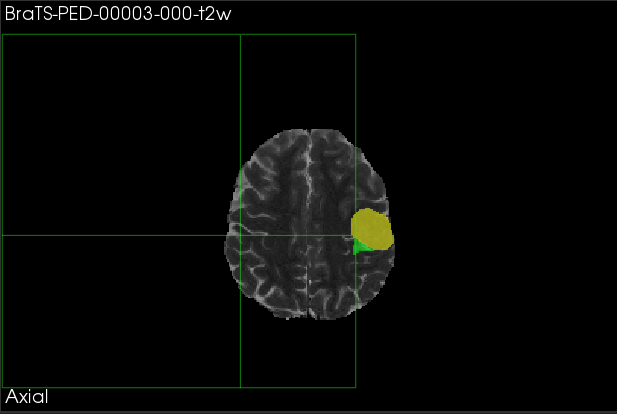}&
        \includegraphics[width=2.5cm]{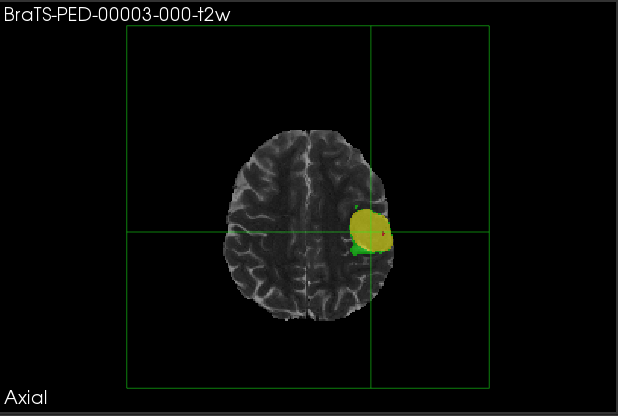} \\
    
        (j) &
        \includegraphics[width=2.5cm]{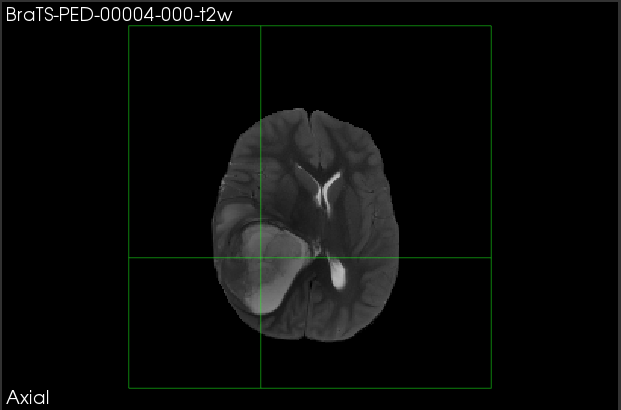} &
        \includegraphics[width=2.5cm]{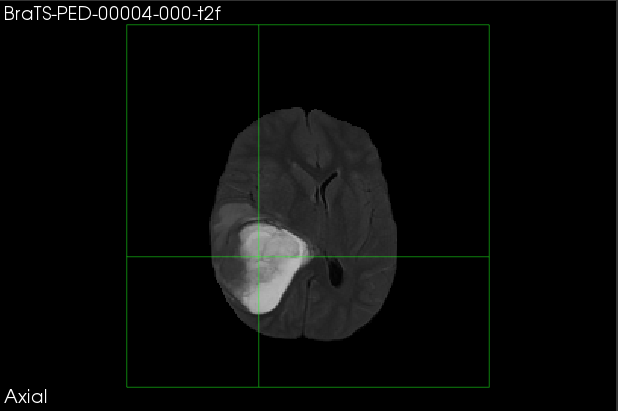} &
        \includegraphics[width=2.5cm]{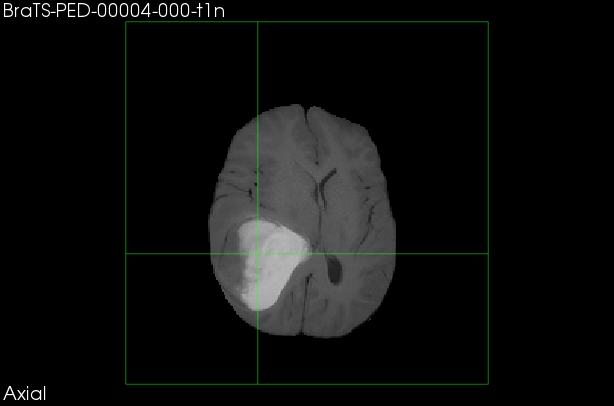} &
        \includegraphics[width=2.5cm]{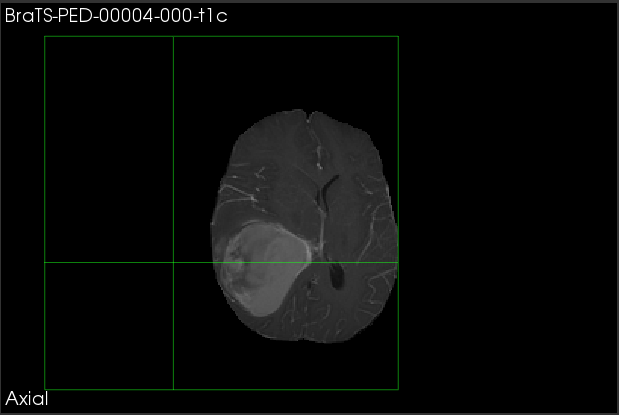}&
        \includegraphics[width=2.5cm]{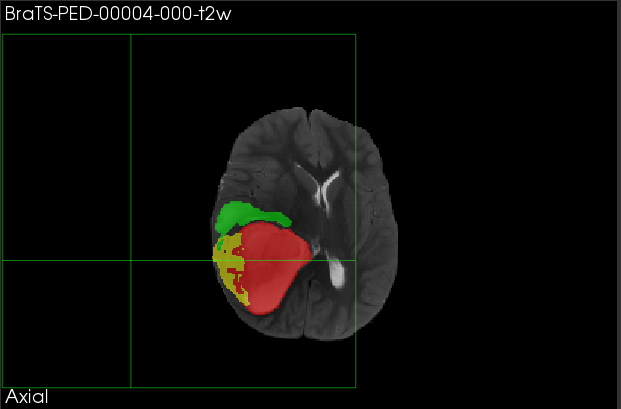}&
        \includegraphics[width=2.5cm]{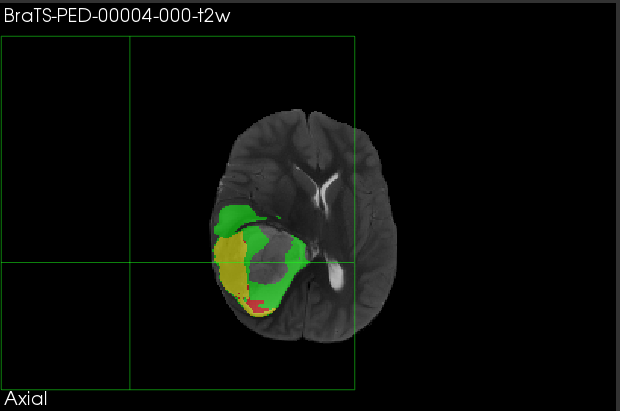} \\

    \end{tabular}
    \caption{This table displays input data, which comprises T2W, T2F, T1N, and T1C images. We compare these input images with corresponding ground truth data and our predictions. Specifically, (a, b) represent data for Adult Glioma, (c, d) for BraTS-Africa, (e, f) for Meningioma, (g, h) for Brain Metastases, and (i, j) for Pediatric Tumor.}
    \label{compare}
\end{table}

            \section{Discussion}

        Across both the Brain Metastases Dataset and the BraTS-Africa Dataset datasets, the results highlight the effectiveness of the proposed multi-modal attention-based tumor segmentation model, particularly when leveraging multiscale inputs and Omni-dimensional Dynamic Convolution. The combination of these approaches improves the segmentation accuracy and performance of lesion subregions. The data processing combination further refines results. These findings contribute to more accurate and reliable tumor segmentation.

    \section{Conclusion}

        This paper has provided an in-depth exploration of our proposed model's performance across distinct datasets. The combined utilization of multiscale inputs, ODConv techniques, and strategic data processing leads to enhanced segmentation accuracy, showcasing the potential for more accurate and reliable tumor segmentation across various medical imaging scenarios.
    
    % =========================
    \iffalse
    \section*{Author Contributions}
        Study conception and design: S.K.M., V.J., U.B., S.C.G.
        Software development used in the study: S.K.M., A.G., S.S., A.G.
        Wrote the paper: S.K.M., A.G., S.R., V.J., U.B., S.C.G.
        Data analysis and interpretation: S.K.M., V.J., U.B., S.C.G.
        Reviewed/edited the paper: S.K.M., S.R., V.J., U.B., S.C.G.
    
    \section*{Acknowledgments}
        The success of any challenge in the medical imaging domain depends upon the quality of well-annotated multi-institutional datasets. We thank all the data contributors, annotators, and approvers for their time and efforts.
         %This section should be added at the conclusion of the challenge, to avoid revealing data sources during the challenge. The data contributors are: 1)Center 1  2) Center 2 
    \fi

    \bibliographystyle{ieeetr}
    \bibliography{bibliography.bib}
    \newpage
    \appendix
\end{document}